\documentclass[pdflatex,sn-mathphys-num]{sn-jnl}

\usepackage{graphicx}%
\usepackage{multirow}%
\usepackage{amsmath,amssymb,amsfonts}%
\usepackage{amsthm}%
\usepackage{mathrsfs}%
\usepackage{xcolor}%
\usepackage{textcomp}%
\usepackage{manyfoot}%
\usepackage{booktabs}%
\usepackage{algorithm}%
\usepackage{algorithmicx}%
\usepackage{algpseudocode}%
\usepackage{listings}%
\usepackage[caption=false]{subfig}
\usepackage{tabularx}
\usepackage{float}
\usepackage{hyperref}
\usepackage[utf8]{inputenc}
\usepackage[T1]{fontenc}
\usepackage{CJKutf8}
\usepackage{CJKfntef}
\usepackage{makecell}
\usepackage{tcolorbox}
\usepackage{pifont}
\usepackage{geometry}
\usepackage{lmodern}
\usepackage{anyfontsize}
\usepackage{enumitem}

\begin{document}

\title[Article Title]{SynAsk: Unleashing the Power of Large Language Models in Organic Synthesis}


\author[1]{\fnm{Chonghuan} \sur{Zhang}}
\equalcont{These authors contributed equally to this work.}

\author[1]{\fnm{Qianghua} \sur{Lin}}
\equalcont{These authors contributed equally to this work.}

\author[2]{\fnm{Biwei} \sur{Zhu}}
\equalcont{These authors contributed equally to this work.}

\author[2]{\fnm{Haopeng} \sur{Yang}}

\author[2]{\fnm{Xiao} \sur{Lian}}

\author[2]{\fnm{Hao} \sur{Deng}}%

\author[2]{\fnm{Jiajun} \sur{Zheng}}%

\author*[1]{\fnm{Kuangbiao} \sur{Liao}}\email{liao\textunderscore kuangbiao@gzlab.ac.cn}

\affil[1]{\orgname{Guangzhou National Laboratory},  \orgaddress{\city{Guangzhou}, \state{Guangdong}, \country{PR China}, \postcode{510005}}}

\affil[2]{\orgname{AIChemEco Inc.}, \orgaddress{\city{Guangzhou}, \state{Guangdong}, \country{PR China}, \postcode{510005}}}


\abstract{The field of natural language processing (NLP) has witnessed a transformative shift with the emergence of large language models (LLMs), revolutionizing various language tasks and applications, and the integration of LLM into specialized domains enhances their capabilities for domain-specific applications. Notably, NLP has made significant strides in organic chemistry, particularly in predicting synthetic tasks, paving the way for the development of LLMs tailored to the organic chemistry field. In this work, we introduce SynAsk, a comprehensive organic chemistry domain-specific LLM platform developed by AIChemEco Inc. By finetuning an LLM with domain-specific data and integrating it with a chain of thought approach, SynAsk seamlessly accesses our knowledge base and advanced chemistry tools in a question-and-answer format. This includes functionalities such as a basic chemistry knowledge base, molecular information retrieval, reaction performance prediction, retrosynthesis prediction, chemical literature acquisition, and more. This novel methodology synergizes fine-tuning techniques with external resource integration, resulting in an organic chemistry-specific model poised to facilitate research and discovery in the field. Accessible via \url{https://synask.aichemeco.com}, SynAsk represents a significant advancement in leveraging NLP for synthetic applications.}

\keywords{Large Language Model, AI in Chemistry, organic synthesis, retrosynthesis}



\maketitle

\section{Introduction}\label{sec1}

In recent years, the field of natural language processing (NLP) has undergone a revolutionary shift with the emergence of large language models (LLMs), advanced artificial intelligence systems trained on massive datasets to understand and generate human-like text across various language tasks and applications. At the core of LLMs lies the remarkable technology of generative pre-trained transformers (GPT) \cite{gpt}. Developed by OpenAI, GPT models like ChatGPT \cite{chatgpt} have gained widespread attention and adoption for their capacity to produce coherent and contextually relevant text. ChatGPT, in particular, represents a milestone in conversational AI, enabling human-like interactions that go beyond scripted responses. Evolving from ChatGPT to GPT-4 \cite{gpt4} through continual learning from vast datasets allows these models to grasp nuances of language and context, making them versatile tools for diverse tasks, from assisting in creative writing to generating videos. While GPT models have dominated the landscape, other models like Qwen \cite{qwen} and LLaMA \cite{llama} also make significant contributions to the field, and these models are open-sourced for the community to utilize. Qwen, primarily trained from Mandarin Chinese language sources, is renowned for its robustness in question-answering tasks, leveraging a different architecture and training approach. On the other hand, LLaMA specializes in language understanding and inference tasks, offering unique capabilities in semantic analysis and knowledge extraction.

Beyond ChatGPT and other models, LLMs encompass a spectrum of applications across vertical domains. Domain-specific and customized data have been collected and labeled to fine-tune these LLMs. One of the key benefits of vertically specialized LLMs is their capacity to bolster domain-specific applications. By refining their expertise within a particular domain, these models possess the capability to delve deeply into the nuances of the subject matter, rendering them invaluable tools for professionals operating in specialized domains. For instance, a legally specialized LLM, namely DISC-LawLLM \cite{lawllm}, can provide precise legal counsel, draft contracts, and facilitate intricate legal research, thereby streamlining processes and conserving resources for legal practitioners. Similarly, a medically specialized LLM, namely MultiMedQA \cite{medllm}, can assist physicians in diagnosing rare conditions, proposing tailored treatment plan, and staying updated on the latest technologies in medical research.

The integration of NLP into organic chemistry has brought about a revolution in research and discovery. Molecules and reactions can now be represented using SMILES (Simplified Molecular Input Line Entry System), a textual notation for depicting high-dimensional chemical structures \cite{smiles}. NLP techniques have been employed to tackle organic synthesis tasks using SMILES strings, treating the synthesis problem as a sequence generation task. This approach involves training machine learning models to predict the sequence of molecules and reactions necessary to synthesize a target molecule based on desired products. These models learn from extensive datasets of annotated reactions, where each reaction is represented as a sequence of SMILES strings. Leveraging the patterns and rules encoded in the data, these models can generate plausible synthesis pathways \cite{moltransformer, cdi}. 

LLMs have found applications in organic chemistry as well. However, without further tuning with organic chemistry domain-specific data, researchers have evaluated five LLMs in tasks related to organic chemistry, including reaction prediction and retrosynthesis. While these models provide reasonable results in classification or ranking tasks like yield prediction and reagent selection, they face challenges in generative tasks that require a deep understanding of molecular structures \cite{llm_in_chem}. This difficulty may stem from the highly experimental nature of organic chemistry, the lack of labeled data, and the limited scope and applicability of computational tools in this field \cite{chemcrow}. To bridge this gap and motivate further exploration of LLM potential in chemistry, several domain-specific LLMs for organic chemistry have been developed. ChemCrow \cite{chemcrow} was the first proposed LLM in chemistry aimed at enhancing its capabilities through external tools. It employs chain-of-thought (CoT) strategies \cite{wei2022chain}, which are a series of intermediate reasoning steps to improve LLMs' ability to understand tasks from prompts. ChemCrow also utilizes LangChain\cite{topsakal2023creating}, a framework to connect the LLM with multiple external tools downstream to solve specific tasks and return answers back to the LLM. However, this method relies on the reliability of tools, and general LLMs may not comprehensively understand prompts and link to the correct tools to solve specific tasks.Another approach, ChemLLM \cite{chemllm}, was proposed to transform structured chemical data into forms suitable for LLMs to fine-tune the LLaMA model. ChemLLM excels in tasks such as cheminformatic programming. However, its performance may not be as robust as comprehensive models like ChatGPT-4, possibly due to human biases in the collection of incomplete structural chemical data. 

\begin{figure}[ht]
\centering
\includegraphics[width=\textwidth]{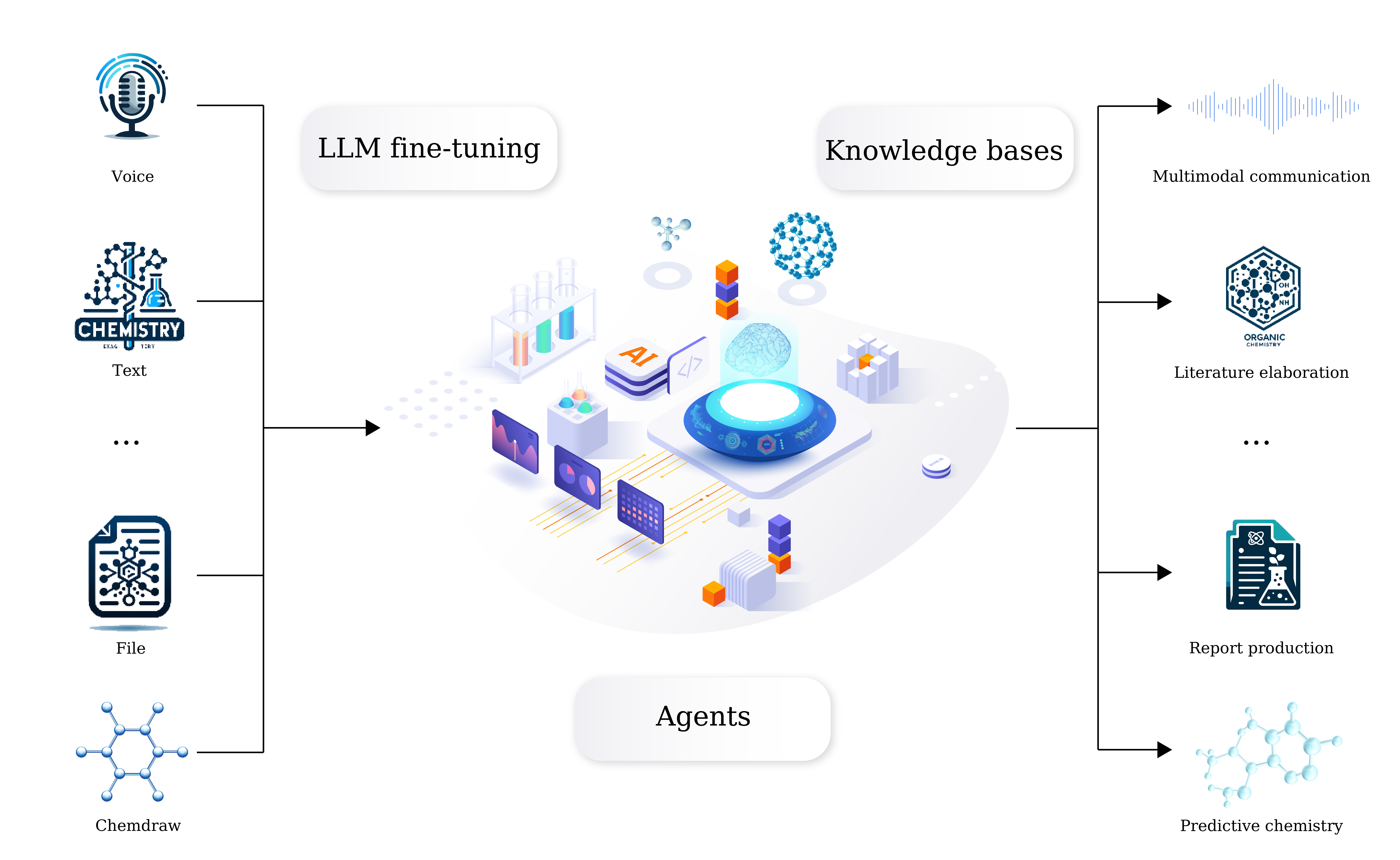}
\caption{\centering The overview of SynAsk platform.}\label{intro_fig}
\end{figure}

We has long been dedicated to AI in chemistry research, developing a series of machine learning and computational based tools to solve fundamental organic chemistry tasks. However, we recognize that directly connecting these tools to large language models (LLMs) may not yield appropriate results. Here we introduce a comprehensive domain-specific LLM for organic chemistry developed by AIChemEco, named SynAsk, as shown in \autoref{intro_fig}. An LLM was refined using a limited set of domain-specific chemistry data and integrate it with a chain-of-thought approach to understand user prompts. Our aim is to utilize Langchain to seamlessly connect SynAsk with our existing suite of tools, addressing specific user inquiries, drawing on the framework of Langchain$-$chatchat \cite{langchain-chatchat}. This methodology allows us to combine fine-tuning techniques with the integration of external resources, resulting in the development of an organic chemistry-specific model. The model can be accessed at \url{https://synask.aichemeco.com}.

\newpage
\section{Methods}\label{sec2}

To construct the comprehensive model integration platform, our approach unfolds along three primary dimensions: utilizing a powerful foundation LLM as the base for SynAsk, crafting more effective prompts and implementing fine-tuning to the foundation model, and connecting with multiple tools to assemble a chemistry domain-specific model platform.

\subsection{Selection of a foundation LLM}
Through various experiments, we have recognized that for the foundation LLM to effectively understand prompts from end-users and apply insights to decide whether to provide LLM inference answers or use specific tools to resolve downstream tasks, it needs to have at least 14 billion parameters. Therefore, only foundation models with over 14 billion parameters were considered. The capabilities of the LLM was assessed using indicators such as Massive Multi-task Language Understanding (MMLU) \cite{hendrycks2020measuring},  Multi-level multi-discipline chinese evaluation (C-Eval) \cite{huang2024c}, GSM8K \cite{cobbe2021training}, (BIG-Bench-Hard) BBH \cite{suzgun2022challenging} and Measuring massive multitask language understanding in Chinese (CMMLU) \cite{li2023cmmlu}, as elaborated in \autoref{esi_indicators} of Electronic Supplementary Information (ESI). These indicators collectively offer a comprehensive assessment of a model's proficiency, covering areas such as linguistic understanding, mathematical reasoning, contextual comprehension, multi-modal integration, and the application of Chain-of-Thought (CoT), which evaluates the fluency of LLMs' integration with external tools. This evaluation framework underscores the essential and diverse skills a model must possess to adeptly address complex real-world problems.

As indicated in \autoref{tab:model_performance} \cite{qwen}, the Qwen series \cite{qwen} outperforms other models with equivalent parameter counts, including LLaMA2 \cite{touvron2023llama}, ChatGLM2 \cite{du2022glm}, InterLM \cite{team2023internlm}, Baichuan2 \cite{yang2023baichuan} and Yi \cite{ai2024yi} in these areas. Additionally, our testing has confirmed that the Qwen series is more compatible with our framework, especially with the release of Qwen-1.5, which provides us with more options. We acknowledge that GPT series \cite{chatgpt}, particularly GPT-4 \cite{gpt4}, scores higher than Qwen. However, at the time of this work, GPT-4 has not been open-sourced and requires paid API tokens to use as a foundation model. To ensure SynAsk remains publicly accessible, we opted to use only open-sourced foundation LLMs and developed an architecture that allows for smooth switching of the foundation LLM, as discussed in \autoref{architecture}.

\subsection{Refinement to more Reasonable Prompt}
To improve the model's performance in two key areas—providing more targeted responses in the chemical domain and enhancing its ability to efficiently utilize tools—we refined our prompt templates through iterative testing and adjustments. We guide the model to generate responses that are not only accurate but also consistent with specific demand expectations. This process encourages the model to become more deeply involved in the task at hand, reducing ambiguity and focusing its attention. These optimized guidance models function as both competent chemists and skilled tool users, establishing a more focused, efficient, and effective interaction between the model and the user.

In our integrated platform, utilizing the classification function of LLMs is particularly crucial, as illustrated in \autoref{fig:synask_workflow}. Since this platform extends from our existing NLP project, we believe it inherently possesses enhanced capabilities. To further train it, we employ a tailored hint project, where the model's role is set as a chemist evaluating and scoring the generated results. This project provides several examples to guide the model. This setup enables the model to discern whether responses augmented by the knowledge database meet the criteria, thereby classifying the results into those that meet expectations and those that do not.

\subsection{Fine-tuning of the LLM}
The selected model underwent fine-tuning to specialize it further in the field of chemistry, ensuring its engagement in professional chemical dialogues, particularly in organic synthesis. The fine-tuning process comprised two iterations, with data processed accordingly for each iteration.
\begin{itemize}
    \item The first iteration was supervised fine-tuning: This stage focused on enhancing the model's cognitive abilities, reinforcing its identity as an expert in chemistry. The objective was to delve deeper into the model's capabilities within the chemistry domain without expanding its original data source. This approach allowed the model to utilize existing data more effectively to solve chemical problems.
    \item The second iteration was instruction-based fine-tuning: The aim here was to improve the model's reasoning and tool invocation capabilities, thereby enhancing its chain of thought. It learned to differentiate between various types of chemical identifiers, such as SMILES and CAS numbers, rather than treating them as ordinary words or sequences of numbers.
\end{itemize}

The rationale for dividing the fine-tuning into two stages is threefold:
\begin{itemize}
    \item Clear and Controllable Training: Each fine-tuning task addressed a specific sub-problem, ensuring clarity and controllability in the training process and outcomes. This approach facilitates adjustments and improvements based on the results of previous fine-tuning, gradually enhancing the model's performance on specific tasks.
    \item Prevention of Interference:  Segregating the tasks prevents confusion and interference between them. Combining all tasks into a single fine-tuning session might lead to instability in training or reduced performance.
    \item Accelerated Training: This approach speeds up the training process. By simplifying each fine-tuning task, the training becomes more efficient, yielding quicker results and feedback. The shorter training times for each task contribute to a faster overall training cycle.
\end{itemize}

After fine-tuning, detailed techniques, procedures, and the necessary equipment are elaborated in \autoref{esi_finetune} of the ESI. Post-fine-tuning, our emphasis mainly lies on the model's ability to demonstrate Chain of Thought (CoT) in its output. Following the fine-tuning process, we provide two examples of the model's simplified output format:
\\[1em]
\noindent
\begin{tabular*}{\textwidth}{@{\extracolsep{\fill}} ll}
\textbf{Prompt:} & What is the SMILES of toluene? \\
\textbf{Response:} & 
\begin{minipage}[t]{.8\textwidth}
\begin{verbatim}
Action: get_SMILES
Action Input: {"query": "toluene"}
\end{verbatim}
\end{minipage} \\
\textbf{Prompt:} & What is the name of CC1=CC=CC=C1? \\
\textbf{Response:} & 
\begin{minipage}[t]{.8\textwidth}
\begin{verbatim}
Action: CAStoName
Action Input: {"query": "CC1=CC=CC=C1"}
\end{verbatim}
\end{minipage} \\
\end{tabular*}
\\[1em]
Notably, the power of these fine-tuned results is significantly enhanced when used in conjunction with appropriately designed prompting strategies and specially designed tool formats. These responses demonstrate the model's ability to identify the required action and its corresponding input from the prompts. However, within our framework, these responses are not the final outcome. Instead, they serve as intermediate prompts to be re-fed into the model. This intermediary step is pivotal, enabling the model to discern the specific tool it requires (e.g., `get\_SMILES' for the initial example) and to process the "Action Input" (e.g., "query: 'toluene'") utilizing the designated tool. Subsequently, the expansive model amalgamates the tool's output with its vast knowledge base, culminating in the generation of a final answer.

\subsection{SynAsk Architecture}\label{architecture}
In the final phase, we implemented the LangChain framework to seamlessly integrate our local knowledge base with both internal and external open-source tools and APIs. Its primary role is to interpret the outputs from the language models, converting them into a format understandable by external tools, thus facilitating the execution of corresponding actions. Simultaneously, it translates the responses from these tools back into a form comprehensible by the language models. Furthermore, LangChain's support for context management enables it to track the interaction history between users and the system. This enhances the system's ability to understand user intentions and maintain session continuity during interactions with external tools. Its scalability ensures that the system can adapt to technological advancements and changing user demands, providing a dynamic and responsive framework for our integration needs. The LangChain framework serves as a pivotal bridge, culminating in a logically coherent and systematically robust integration platform known as SynAsk.

\begin{figure}[ht]
\centering
\includegraphics[width=0.8\textwidth]{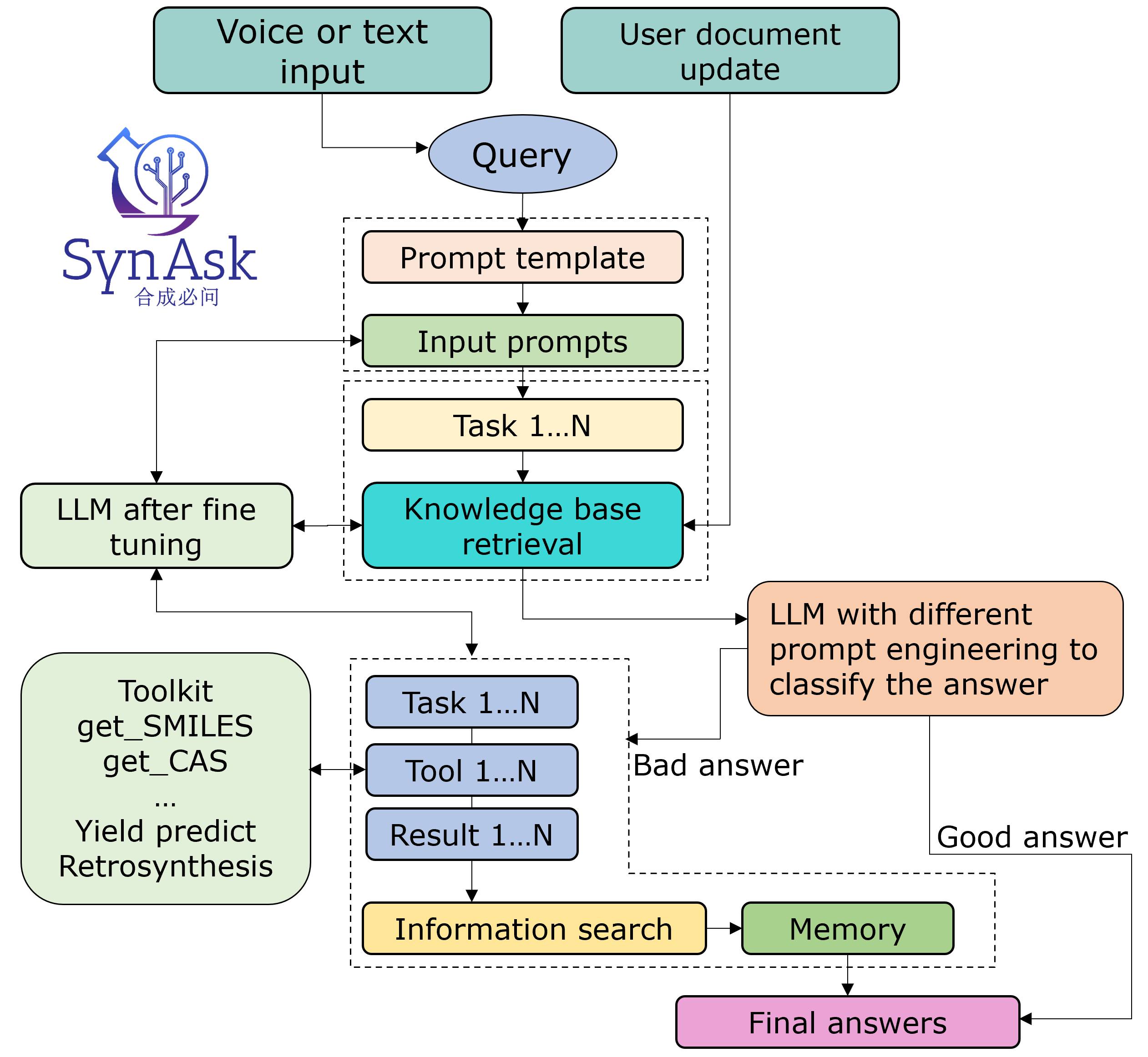}
\caption{\centering The workflow of the SynAsk platform: from the input to the final answer.}\label{fig:synask_workflow}
\end{figure}

The structural framework of SynAsk is illustrated in \autoref{fig:synask_workflow}. Initially, it can accept both voice and text inputs as queries, which are then segmented into multiple tasks by a LLM and matched against our knowledge base. At this stage, users also have the option to upload their local files as supplementary knowledge or directly engage in conversations with the uploaded files. Once matching texts are obtained, the large model synthesizes the content along with its understanding of the question to deduce a conclusion, thereby generating a result. Subsequently, the model evaluates this result to determine if it meets the expected criteria. If the outcome is deemed satisfactory, it is directly outputted as the Final Answer. Conversely, if the results do not meet expectations, we will enter our customized Agent Q\&A mode and call our tools to answer. Finally, the tool output is combined with the LLM's self-knowledge to generate the final answer.

In the SynAsk architecture, although we currently utilize Qwen-1.5 as the foundation LLM, we recognize the ongoing revolutions in LLM technology. Consequently, we have developed a workflow to swiftly adjust the foundation model and fine-tune the domain-specific data. This approach ensures that SynAsk can continuously update and iterate, leveraging the latest advancements in foundation LLMs.

\subsection{SynAsk Toolsets}
Chemoinformatic tools are seamlessly connected with SynAsk through LangChain to provide comprehensive organic synthesis answers. This includes a variety of machine learning-powered tools developed both internally and by external teams, all dedicated to solving organic synthesis tasks. At the time of publishing this work, \textbf{12} internal tools and \textbf{10} external tools have been integrated into SynAsk. External tools are appropriately cited with their origins. With the rapid development of this field, we anticipate an increasing influx of tools joining SynAsk. These tools are categorized into molecular, reaction tools, and others, with a number of advanced in-house tools elaborated in \autoref{inhouse_tool}.

\subsubsection{Molecular Information Retrieval}
This category encompasses tools designed for querying various molecular identifiers and properties. Functions include retrieving Chemical Abstracts Service (CAS) numbers, Simplified Molecular Input Line Entry System (SMILES) strings, molecular weights, assessing molecular similarity, identifying types of functional groups, and checking the regulatory status of molecules. The respective tools for these purposes are:

\begin{itemize}
    \item \texttt{get\_cas} -- for CAS numbers retrieval \cite{kim2023pubchem}
    \item \texttt{get\_smiles} -- for obtaining SMILES strings \cite{kim2023pubchem}
    \item \texttt{CAStoName} -- to convert CAS numbers to chemical names \cite{pence2010chemspider}
    \item \texttt{SMILEStoName} -- to convert SMILES strings to chemical names\cite{pence2010chemspider}
    \item \texttt{get\_mol\_weight} -- for calculating molecular weights
    \item \texttt{get\_mol\_similarity} -- to determine molecular similarity
    \item \texttt{check\_functional\_groups} -- for functional group identification
    \item \texttt{ControlmolCheck} -- to check if a molecule is controlled
\end{itemize}

\subsubsection{Chemical Reaction and Retrosynthesis}
This category aids in querying chemical reaction conditions, planning chemical reaction pathways, predicting chemical reaction yields, performing retrosynthetic analysis, and predicting reaction derivatives. Tools provided for these functions include:

\begin{itemize}
    \item \texttt{Get\_condition} -- to query chemical reaction conditions
    \item \texttt{ReactionPlanner} -- for planning chemical reaction pathways \cite{rxn4chemistry2023}
    \item \texttt{ReagentsPredict} -- to predict reagents in chemical reactions
    \item \texttt{YieldPredict} -- for predicting chemical reaction yields
    \item \texttt{Retrosynthesis} -- to perform retrosynthetic analysis
    \item \texttt{DerivatePredict} -- to predicts the derivatives from a chemical reaction, using reactants' names or SMILES, enhancing the exploration of reaction outcomes.
    \item \texttt{AutoMapping} -- to identify the position of each atom in the molecules before and after a chemical reaction \cite{schwaller2020unsupervised}\cite{chen2024precise}
\end{itemize}

\subsubsection{Acquisition of Chemical Literature and Knowledge}
Dedicated to acquiring chemical literature and extracting chemical knowledge, tools in this section include:

\begin{itemize}
    \item \texttt{Get\_literature} -- for retrieving literature \cite{ginsparg2011arxiv}\cite{serpapi2023}
    \item \texttt{get\_knowledge} -- to obtain chemical knowledge \cite{serpapi2023}
    \item \texttt{Rxn\_literature} -- for sourcing reaction-specific literature
\end{itemize}

\subsubsection{Miscellaneous}
This section covers a diverse array of functions including drawing chemical molecular structures and balancing chemical equations. Tools include:

\begin{itemize}
    \item \texttt{Moldraw} -- for drawing chemical molecular structures 
    \item \texttt{calculate} -- a general-purpose calculation tool
    \item \texttt{automatic\_balance} -- to automatically balance chemical equations \cite{dahlgren2018chempy}
    \item \texttt{image\_gen} -- for generating and searching images \cite{serpapi2023}
\end{itemize}

\subsubsection{Advanced In-House Analytical Tools}\label{inhouse_tool}
\begin{figure}[H]
\centering
\includegraphics[width=\textwidth]{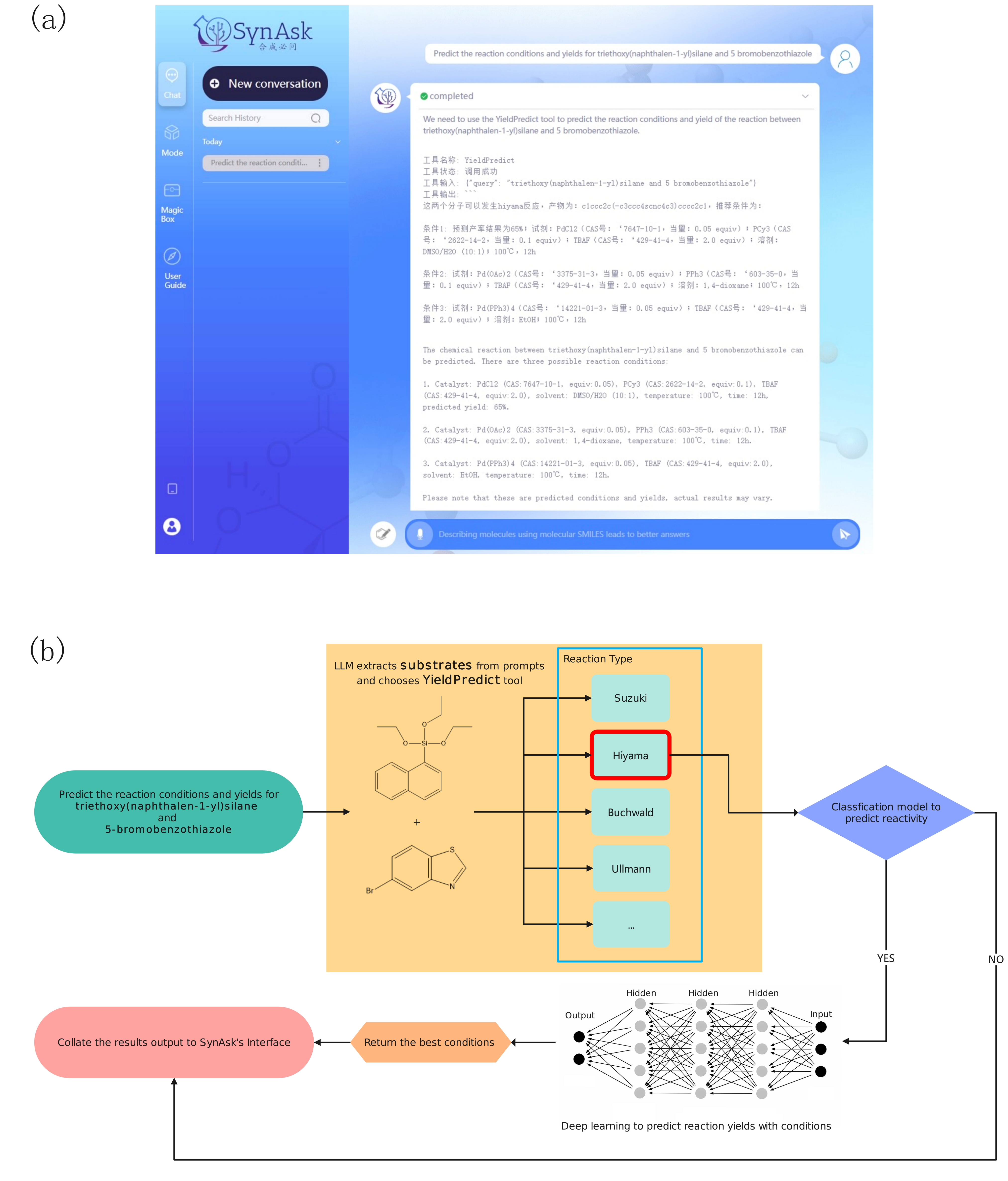}
\caption{\centering An example of the YieldPredict tool workflow for predicting the reaction yield of triethoxy(naphthalen-1-yl)silane and 5-bromobenzothiazole: (a) the user interface of SynAsk, (b) the thinking process of the YieldPredict tool.}\label{yield_pred_example}
\end{figure}

\paragraph{YieldPredict} 
This is an API tool linked with our self-developed reaction yield prediction tool. By inputting at least two substrates, either in their molecular name or molecular SMILES, this tool can identify the possible reaction types of the molecules by querying our reaction template library. With the known reaction types, the molecules are passed into the reaction models as substrates. The models then suggest products and the most suitable reaction reagents and conditions for the substrates. For example, by asking the reaction yield of triethoxy(naphthalen-1-yl)silane and 5-bromobenzothiazole, the tool first parses the two molecules into the reaction templates as substrates \autoref{yield_pred_example}. This suggests Hiyama cross-coupling reactions. The two substrates are then parsed into the Hiyama reaction models, generating products and possible reaction yields under specific reaction reagents and conditions.

We have dedicated our efforts to developing data-driven reaction yield prediction models for common reaction types \cite{ylides, alcohol_oxidation, oh_insertion, dhp}. For each model of a specific reaction type, we conduct chemical reaction experiments using high-throughput experimentation (HTE) techniques with various substrates. This enables us to draw insights from existing literature data and identify areas where experimental data collection is necessary to augment an equitable data space for refining model training, thus facilitating more robust interpolation. We develop reaction models using machine learning techniques such as support vector machine (SVM) and NLP deep learning models like BERT (Bidirectional Encoder Representations from Transformers) \cite{BERT}. These models are validated using external literature test data, achieving reasonable Mean Absolute Error (MAE), commonly below 0.15. As of the publication of this work, we have included 18 reaction types in this tool.

\paragraph{Get\_Conditions}
This tool is a simplified version of YieldPredict. Instead of predicting the reaction product and yield, it provides rapid responses and suggests only the suitable reaction conditions and reagents for the substrates.

\paragraph{Retrosynthesis}
By inputting the desired target products, this tool generates numerous reaction pathways of molecules starting from buyable precursors. We have developed our own retrosynthesis model for this purpose. For a desired product, it is parsed into the reaction template library to find possible substrates and, consequently, the suitable reaction site for bond breakage. A reinforcement learning-trained agent selects the most suitable reaction from the candidates based on the forecasted synthesis difficulty and predicted reaction yield of the substrates (desired products at the previous step). This process is conducted recursively until the last substrates are buyable. At the output, we present the results in both textual form and as retrosynthetic route images. The algorithm of our retrosynthesis model will be published elsewhere.

\newpage
\section{SynAsk Performance}
We evaluate the performance of SynAsk from two perspectives: its general ability as a large language model (LLM), and its proficiency in synthetic chemistry. Additionally, we provide several examples of SynAsk's outputs to demonstrate the platform's comprehension capabilities.

\subsection{General ability of SynAsk}

\begin{figure}[ht]
\centering
\includegraphics[width=0.85\textwidth]{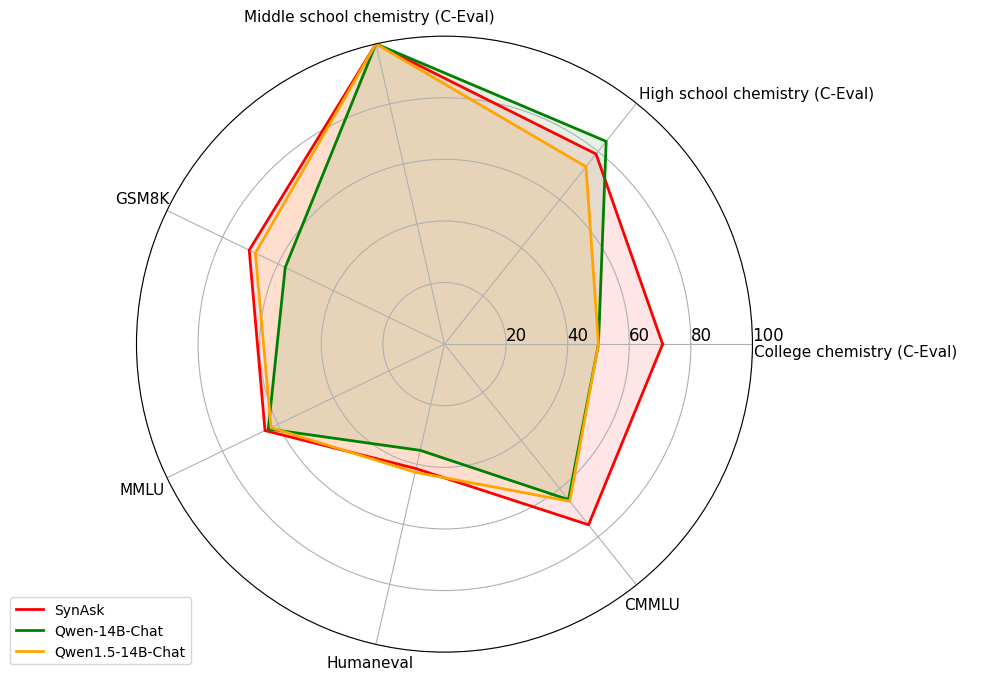}
\caption{\centering The comparison of general ability between SynAsk and Qwen in seven aspects, including its applications in chemistry.}\label{fig:radar}
\end{figure}

We evaluate the performance enhancements achieved through our first fine-tuning method on the SynAsk model based on OpenCompass \cite{2023opencompass}, which serves as a universal evaluation platform for foundation LLMs. The efficacy of the method is demonstrated by its superior scores across various assessment indicators, particularly in its application to chemistry. The definitions of the general indicators used in \autoref{fig:radar} are provided in \autoref{esi_indicators} of the ESI, while the chemistry-related indicators are outlined in \autoref{esi_compass} along with examples. It's noteworthy that indicators such as College Chemistry, High School Chemistry, and Middle School Chemistry in the figure all stem from C-Eval.. SynAsk significantly outperforms its foundation model predecessors. For example, in the area of College Chemistry, SynAsk achieves a remarkable score of 70.83\%, compared to 50\% by both Qwen-14B-Chat and Qwen1.5-14B-Chat. This signifies a substantial improvement, highlighting the model's enhanced ability to effectively utilize existing data sources for solving complex chemical problems..

Furthermore, the scores in other key benchmarks such as MMLU, GSM8K and CMMLU also reflect the overall enhancement of the SynAsk model. In CMMLU, which assesses cross-model multitask learning, SynAsk scored 75.03\%, indicating its proficiency in integrating textual and visual information, crucial for multi-model applications. Similarly, its performance in MMLU and GSM8K benchmarks demonstrate its improved global knowledge comprehension and multi-step mathematical reasoning, respectively.

The advancements in SynAsk are attributed to the fine-tuning approach that leverages existing data sources more efficiently, thus enhancing the model's ability to address nuanced chemical contexts and complex reasoning tasks. This is particularly crucial for applications requiring deep understanding and contextual awareness, as indicated by the improvements in C-Eval scores.

These results collectively underscore the effectiveness of our fine-tuning methodology, confirming its potential to significantly boost performance across diverse linguistic and cognitive challenges, thereby reinforcing the model's utility in academic and practical applications.

\subsection{Proficiency in synthetic chemistry}
The primary proficiency of SynAsk in synthetic chemistry lies in its ability to predict reaction performance, such as yield, and to conduct retrosynthetic planning of target molecules, utilizing the embedded tools within SynAsk. Several case studies are presented and compared with benchmarks to evaluate the model's performance. Additionally, the other functions of SynAsk are compared with ChatGPT-4.0 answers to highlight its advancements in various areas. 

\subsubsection{Reaction yield prediction}
A number of reaction yield prediction models have been developed and widely used to forecast the performance of reactions for frequently encountered reaction classes. For instance, Doyle \textit{et al.}'s palladium-catalysed Buchwald–Hartwig cross-coupling reaction model \cite{doyle} and Richardson \textit{et al.}'s Suzuki-Miyaura cross-coupling reaction model \cite{Richardson} are among the notable examples. These models were trained using self-developed high-throughput experimentation (HTE) reaction data employing machine learning algorithms. Schwaller \textit{et al.} \cite{BERT_yield} further enhanced the performance of these models using the same datasets through pre-trained BERT model. While these methods effectively predict the product yield within the self-developed HTE reaction dataset, their applicability to predicting the product yield of external literature recorded reactions may be limited.

\begin{figure}[ht]
\centering
\includegraphics[width=0.9\textwidth]{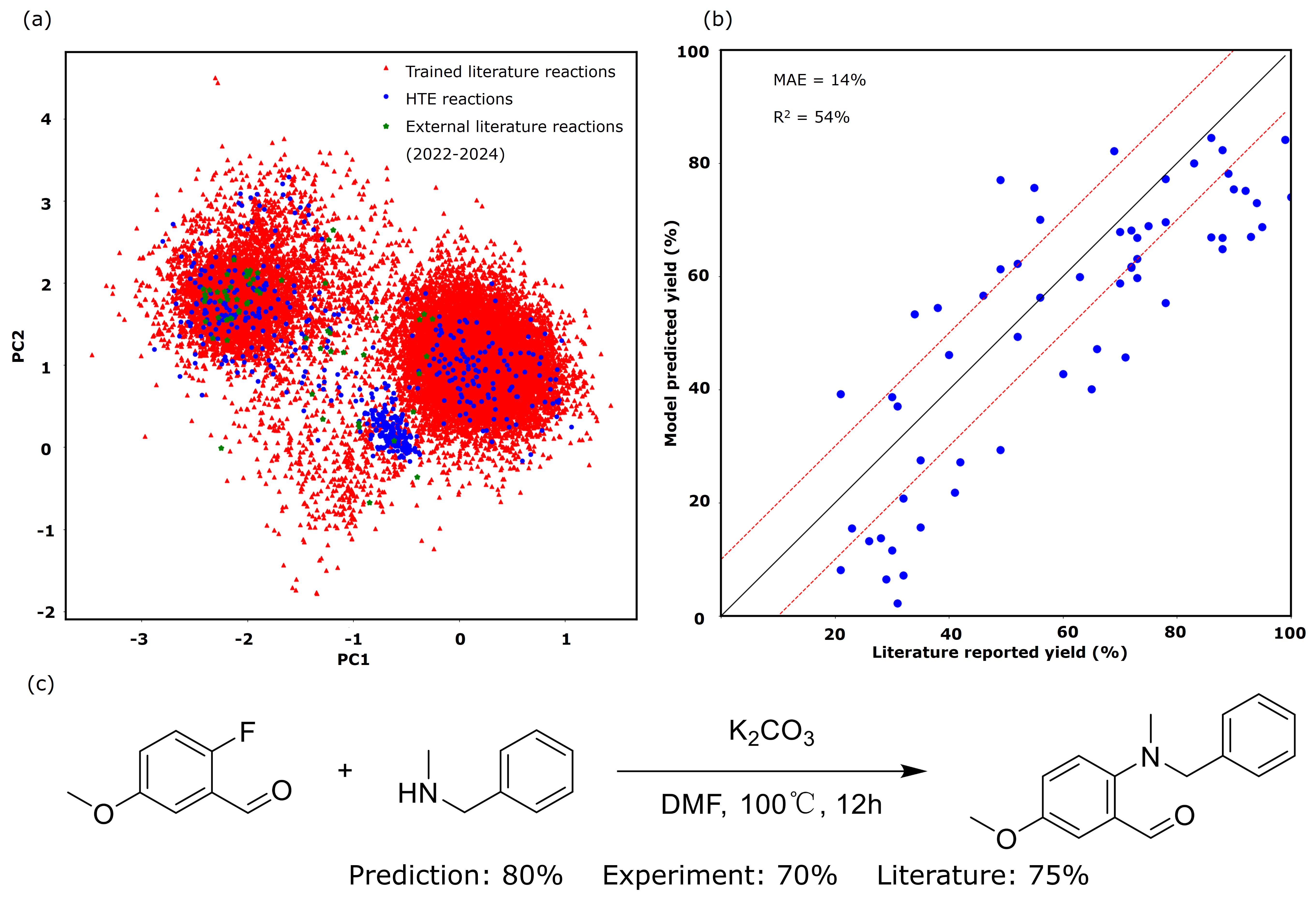}
\caption{\centering The S\textsubscript{N}Ar reaction model results: (a) the chemical space of S\textsubscript{N}Ar reactions under the HTE and literature recorded datasets, (b) the predicted yield versus experimental yield of the test dataset from the three different models, and (c) an example of S\textsubscript{N}Ar reaction: N-methyl-1-phenylmethanamine reacting with 2-fluoro-5-methoxybenzaldehyde.}\label{yield_pred_results}
\end{figure}

We tested our in-house nucleophilic aromatic substitution (S\textsubscript{N}Ar) reaction model embedded in SynAsk with tboth a test set and external literature reaction data. The model test set comprises unseen HTE reaction data, yielding a mean absolute error (MAE) of 11.5\%. For the external literature reaction data, to minimize bias, we randomly collected 60 recently published S\textsubscript{N}Ar reactions from the last three years (2022-2024), including new substrate molecules never seen by the reaction model. The comparison between the model-predicted yield and literature-reported yield is presented in \autoref{yield_pred_results}b, yielding an MAE of 14.1\%. These recent published reactions encompass seven different reaction conditions. For example, N-methyl-1-phenylmethanamine reacting with 2-fluoro-5-methoxybenzaldehyde under K\textsubscript{2}CO\textsubscript{3} and DMF is illustrated in \autoref{yield_pred_results}c. The literature-reported yield of the product 2-(benzyl(methyl)amino)-5-methoxybenzaldehyde is 75\% \cite{yield_example}, whilst our model predicts 80\% and our HTE experimental yield is 70\%. Additional results are provided in \autoref{esi_rxn_pred} of the ESI.

The decay in prediction accuracy observed when transitioning from HTE reactions to literature-reported reactions is primarily attributed to the increased complexity of substrates in literature reactions. These substrates are often more intricate and unseen by the model, thereby encompassing a wider range within the chemical space, as depicted in \autoref{yield_pred_results}a. To compute the chemical space, we digitized the reactions using RXNFP pretrained reaction fingerprint \cite{rxnfp} and reduced into two dimensions. \autoref{yield_pred_results}a also weakly show three clusters of the S\textsubscript{N}Ar reaction. Despite the challenges posed by the complexity of literature-reported reactions, our in-house S\textsubscript{N}Ar model demonstrates the capability to predict the reaction performance of these reactions. This is particularly valuable as it enables predictions closer to the reactions of interest to synthetic chemists.

\subsubsection{Retrosynthetic route planning}
We tasked SynAsk with planning retrosynthetic routes for 11,549 small molecule drugs recorded in the ChEMBL database \cite{chembl_drugs}. SynAsk successfully predicted retrosynthetic routes for 6,358 molecules, suggesting step-by-step routes starting from buyable precursors. This accounts for 55\% of the queried molecules. In contrast, the State of the Art (SOTA) open-sourced retrosynthetic planning tool, AIZynthFinder \cite{aizynthfinder}, only suggested 3,118 retrosynthetic routes, covering 27\% of the queried molecules. 

\begin{figure}[H]
\centering
\includegraphics[width=1\textwidth]
{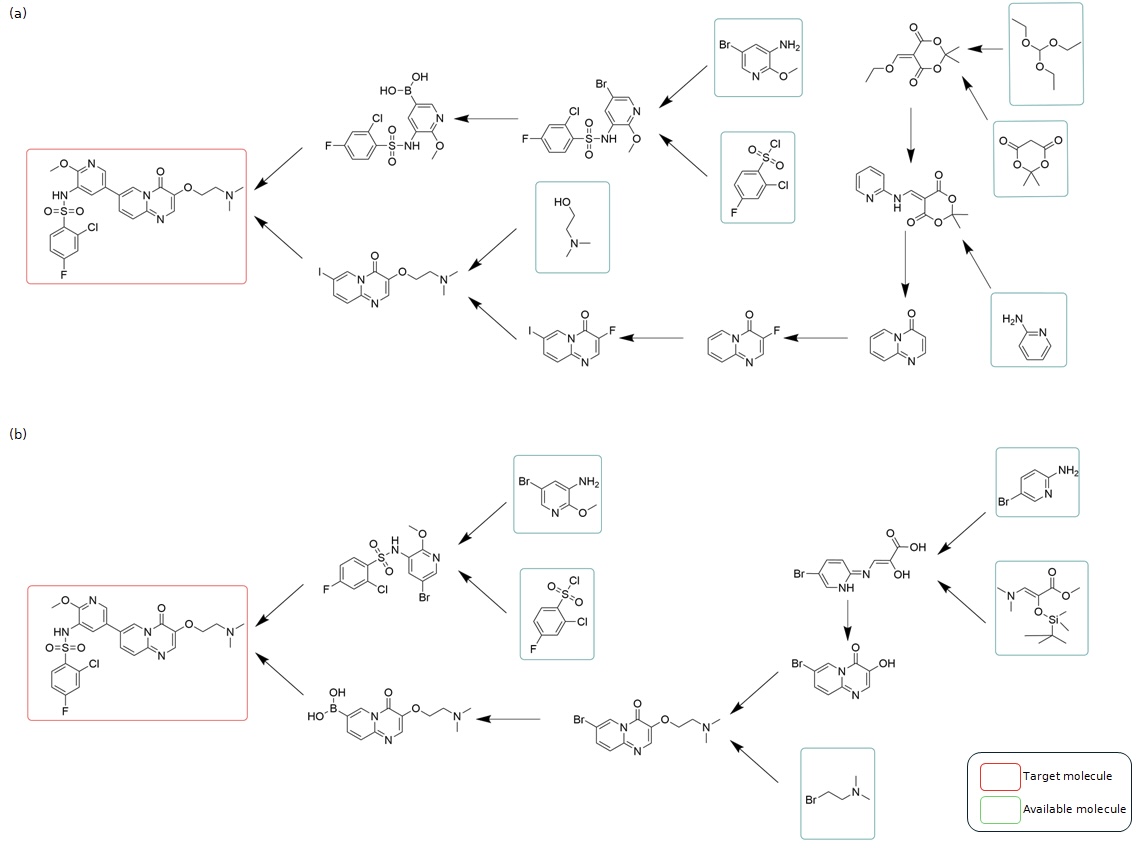}
\caption{\centering The comparison among synthetic routes of the target molecule Gilmelisib: planned by (a) SynAsk's retrosynthetic tool and (b) an experienced synthetic chemist.}\label{retro_example}
\end{figure}

As a case study, let's consider the retrosynthesis of Gilmelisib, a novel small molecule under investigation as a selective inhibitor of PIK3C$\alpha$, potentially treating cancers characterized by PIK3C$\alpha$ mutations. SynAsk proposes a seven-step synthetic route with four precursors (as shown in  \autoref{retro_example}a), matching the route suggested by an experienced human chemist in terms of length and number of precursors (as shown in \autoref{retro_example}b). SynAsk utilizes inexpensive precursors to rapidly obtain key heterocyclic fused ring intermediates through straightforward Knoevenagel condensation and addition-elimination reactions. In contrast, AIZynthFinder fails to provide a synthesis route for the target molecule, even after enriching its starting materials with our lists of buyable precursors. Additional synthetic routes for small molecule drugs are detailed in \autoref{esi_retrosynthesis} of the ESI.

While we refrain from concluding that SynAsk is smarter or approaching the intelligence of a human chemist in retrosynthesis, as this determination would necessitate passing the Turing test \cite{mikulak_retro, segler_retro} or experimental validation, we acknowledge that SynAsk's retrosynthetic ability offers valuable insights for synthetic chemists and assists in synthesis planning.

\subsection{Examples of the SynAsk platform outputs vesus other LLMs}

Here we present a comparative analysis of the performance of three LLMs – SynAsk, ChatGPT-4.0, and ChemCrow – in addressing synthetic chemistry queries. We evaluated their capabilities by inputting a set of synthetic questions, encompassing both general and professional inquiries, to assess their aptitude in providing accurate and relevant responses.

\subsubsection{General inquiries}
Queries such as ``Can you recommend me some reaction conditions for Suzuki cross-coupling?" or ``Please help me find some literature related to C-H activation" were presented to all three LLMs. Across the board, each model exhibited proficiency in generating appropriate responses, showcasing their utility in aiding chemists with routine inquiries, details in \autoref{esi_llm_example} of the ESI.

\subsubsection{Professional synthetic questions}

A more rigorous evaluation was conducted by inputting a specific synthetic question: ``Tell me what reaction can occur between Nc1ccc2nccnc2c1.O=C(O)Cc1cc(F)cc(F)c1 and what the product is." Here ``Nc1ccc2nccnc2c1.O=C(O)Cc1cc(F)cc(F)c1" represents the SMILES syntax for quinoxalin-6-amine and 3,5-Difluorophenylacetic acid as substrates. The deliberate use of SMILES allows us to assess the LLMs' ability to recognize molecules from SMILES.

\begin{figure}[ht]
\centering
\includegraphics[width=1\textwidth]{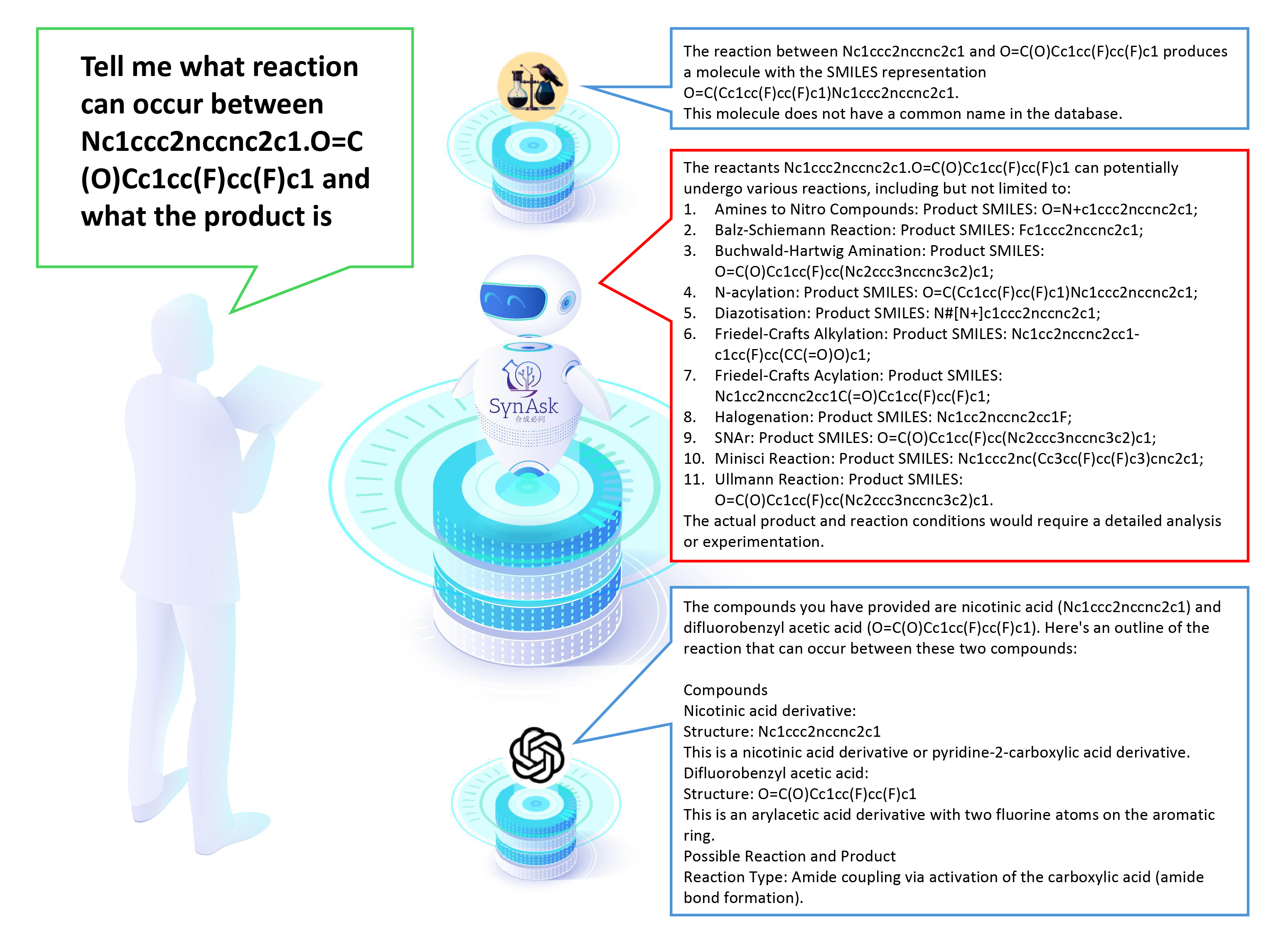}
\caption{\centering The comparison of SynAsk, ChatGPT-4, and ChemCrow output on a professional synthetic question.}\label{compare_output}
\end{figure}

As shown in \autoref{compare_output}, SynAsk demonstrates its specialization in organic chemistry by providing a comprehensive list of potential reactions and their corresponding products. Leveraging its domain-specific knowledge, SynAsk offers a diverse array of feasible transformations, including N-acylation, Buchwald-Hartwig amination, Minisci reaction, among others. This exhaustive output underscores SynAsk's capacity to analyze complex molecular interactions and propose multiple viable pathways.

In contrast, ChemCrow delivers a singular response, identifying the reaction as N-acylation and providing the corresponding product. While ChemCrow offers a concise solution, its limitation in providing alternative reaction pathways restricts its utility in scenarios where multiple transformation possibilities exist.

ChatGPT-4, although proficient in understanding the query, encounters a misinterpretation in identifying the compounds involved. While it accurately delineates the structure and classification of the provided molecules, it erroneously labels Nc1ccc2nccnc2c1 as nicotinic acid derivative, instead of recognizing it as quinoxalin-6-amine. This discrepancy underscores the model's susceptibility to misinterpretation of chemical structures, particularly in complex contexts.

SynAsk distinguishes itself as a specialized LLM tailored specifically for organic chemistry tasks. Its domain-specific training and integration of fine-tuning techniques result in a robust model capable of providing detailed insights and accurate predictions for complex synthetic queries. While ChatGPT-4 and ChemCrow offer general language processing capabilities, they lack the nuanced understanding and domain expertise exhibited by SynAsk in the context of organic chemistry applications. Therefore, for researchers seeking nuanced insights and comprehensive analyses in organic synthesis, SynAsk stands as a valuable tool for augmenting chemical exploration and discovery.

\section{Conclusions and Future Works}
In this work, we have developed SynAsk, a specialized LLM-powered platform for synthetic chemistry. It represents the first publicly accessible chemistry domain-specific LLM, fine-tuned with selected chemistry data and connected with both in-house and external chemoinformatic tools. Through comparative analyses with foundation LLMs, we have demonstrated SynAsk's proficiency and specialization in synthetic chemistry. Results obtained in reaction yield prediction and retrosynthesis further validate SynAsk's capability in providing valuable chemical insights to synthetic chemists across multiple domains.

Looking ahead, our future endeavors aim to enhance the functionality of SynAsk by empowering the language model and fine-tuning it with additional data for more seamless and appropriate responses. Additionally, we envision SynAsk playing a pivotal role in driving autonomous reaction laboratories \cite{robotic_chemist}. Traditionally, reaction robots have been constrained by written scripts to define their scopes. Recent research has showcased the potential of LLMs to drive robotic chemists effectively \cite{coscientist}. Leveraging SynAsk's capabilities such as retrosynthesis, inference, and programming script writing, we foresee it being instrumental in driving autonomous laboratories, representing the next phase of our fusion of LLM and hardware research.

\section{Acknowledgements}
We are grateful for financial support from Guangzhou National Laboratory, and the National Natural Science Foundation of China (22071249, 22393892). 

\section{Conflict of Interest}
We have a patent application in China with the application number 202410714040.6 titled "A Human-Computer Interaction Method and Electronic Device Based on a Large Language Model".

\bibliography{manuscript}

\pagebreak
\pagenumbering{arabic}
\setcounter{section}{0}
\counterwithin{table}{section}
\counterwithin{figure}{section}

\renewcommand{\thetable}{S\arabic{table}}
\renewcommand{\thefigure}{S\arabic{figure}}
\renewcommand{\thesection}{S\arabic{section}}
\begin{titlepage}
    \centering
    \vspace*{2cm}
    \title{\huge SynAsk: Unleashing the Power of Large Language Models in Organic Synthesis}\\
    \author{\vspace{3em}}
    \textbf{Electronic Supplementary Information} \\
    \vspace{2em}
    Chonghuan Zhang\textsuperscript{1}\textsuperscript{†}, Qianghua Lin\textsuperscript{1}\textsuperscript{†}, Biwei Zhu\textsuperscript{2}\textsuperscript{†}, Haopeng Yang\textsuperscript{2}, Xiao Lian\textsuperscript{2}, Hao Deng\textsuperscript{2}, Jiajun Zheng\textsuperscript{2}, Kuangbiao Liao\textsuperscript{1}\textsuperscript{*} \\
    \vspace{2em}
    \textsuperscript{1}Guangzhou National Laboratory, Guangdong, PR China, 510005\\
    \textsuperscript{2}AIChemEco Inc., Guangdong, PR China, 510005\\
    \vspace{3em}
    \textsuperscript{*} Corresponding author(s). E-mail(s): \href{mailto:kuangbiao\_ liao@gzlab.ac.cn}{kuangbiao\textunderscore liao@gzlab.ac.cn}\\ 
    \textsuperscript{†}These authors contributed equally to this work.\\
    \vspace{3em}
    
\end{titlepage}

\newpage

\section{The indicators used to assess LLMs}\label{esi_indicators}
We evaluated the LLM's capabilities using various metrics including Massive Multi-task Language Understanding (MMLU), Multi-level multi-discipline chinese evaluation (C-Eval), GSM8K, BIG-Bench-Hard (BBH), and Measuring massive multitask language understanding in Chinese (CMMLU). These metrics collectively provide a thorough assessment of a model's proficiency, encompassing linguistic understanding, mathematical reasoning, contextual comprehension, multi-modal integration, and the application of CoT, which examines the fluency of LLMs' integration with external tools. This evaluation framework emphasizes the diverse and essential skills a model needs to effectively tackle complex real-world problems.

\begin{itemize}
    \item \textbf{Massive Multi-task Language Understanding, MMLU} represents a comprehensive and multifaceted initiative that aims to evaluate and enhance the performance of language models across a broad range of linguistic challenges, providing an extensive evaluation of global knowledge and problem-solving abilities.

    \item \textbf{Multi-level Multi-discipline Chinese Evaluation, C-Eval} tests models in scenarios that necessitate an understanding of subtle context, which is crucial for applications involving natural language understanding and generation.
    
    \item \textbf{Grade School Math 8K, GSM8K} is a widely recognized test set designed to assess the mathematical capabilities of language models. It comprises problems that require 2-8 steps of basic mathematical operations to test the models' multi-step mathematical reasoning.
    
    \item \textbf{BIG-Bench-Hard, BBH} evaluates language models' capabilities in applying Chain of Thought to humanistic knowledge. It measures how effectively a model can navigate through complex humanistic concepts and ideas, emphasizing its ability to perform sequential reasoning that mirrors human-like understanding in tasks with cultural and historical depth.
    
    \item \textbf{Measuring massive multitask language understanding in Chinese, CMMLU} is a comprehensive Chinese evaluation benchmark specifically used to evaluate the knowledge and reasoning capabilities of language models in the Chinese context. CMMLU covers 67 topics from basic subjects to advanced professional levels.
    
\end{itemize}

\begin{table}[ht]
\centering
\begin{tabularx}{\textwidth}{|l|X|X|X|X|X|}
\hline
Model & \makecell{MMLU \\ (5-shot)} & \makecell{C-Eval \\ (5-shot)} & \makecell{GSM8K \\ (8-shot)} & \makecell{BBH \\ (3-shot)} & CMMLU \\
\hline
LLaMA2-7B  & 46.8          & 32.5            & 16.7           & 38.2         & 31.8     \\
LLaMA2-13B & 55.0          & 41.4            & 29.6           & 45.6         & 38.4  \\
LLaMA2-32B & 62.6         & -               & 42.2           & 44.1         & -     \\
ChatGLM2-6B & 47.9          & 51.7            & 32.4           & 33.7            & -     \\
InterLM-7B  & 51.0          & 53.4            & 31.2           & 37.0         & 51.8     \\
InterLM-20B & 62.1          & 58.8            & 52.6           & 52.5         & 59.0     \\
Baichuan2-7B& 54.7          & 56.3            & 24.6           & 41.6         & 57.1  \\
Baichuan2-13B& 59.5         & 59.0            & 52.8           & 49.0         & 62.0     \\
Yi-34B      & 76.3          & 81.8            & 67.9           & 66.4         & 85.6     \\
Qwen-1.8B    & 45.3          & 56.1            & 32.3            & 22.3         & 52.1     \\
Qwen-7B     & 58.2          & 63.5            & 51.7           & 45.0         & 62.2     \\
Qwen-14B    & 66.3          & 72.1               & 61.3           & 53.4         & 71.0     \\
Qwen-72B    & 77.4          & 83.3            & 78.9           & 67.7         & 83.6  \\
\hline
\end{tabularx}
\captionsetup{justification=centering}
\caption{\centering Performance of Different Models on Various Benchmarks}
\label{tab:model_performance}
\end{table}

\clearpage

\section{Fine-tuning techniques and procedures}\label{esi_finetune}
In our experiments, we explored two distinct fine-tuning methodologies for LLMs. The first approach involved techniques such as quantization to enable the operation of a 14-billion-parameter model within a 24GB GPU environment. The second approach was direct fine-tuning without additional quantization techniques.

For our experiments, we selected a model with 14 billion parameters. We applied Low-Rank Adaptation (LoRA) by incorporating low-rank matrices into the fully connected layers. The parameter details are presented in Table~\ref{table:model_parameters}.

\begin{table}[ht]
\centering
\begin{tabular}{ccc}
\hline
\textbf{Total Parameters} & \textbf{Trainable Parameters} & \textbf{Percentage of Total} \\
\hline
14,209,134,120 & 41,843,040 & $\approx 0.294\%$ \\
\hline
\end{tabular}
\caption{\centering Parameter quantity of the 14-billion-parameter model}
\label{table:model_parameters}
\end{table}

The fine-tuning with quantization process, conducted on a dataset of 200 entries with a batch size of 2, was completed within an hour. This method is a viable solution for managing large model training on hardware with limited memory without significantly compromising precision.

\begin{table}[ht]
\centering
\begin{tabular}{lcc}
\hline
 & \textbf{Before Quantization} & \textbf{After 4-bit Quantization} \\
\hline
During loading & $14 \times 10^9 \times 4$ bytes & $14 \times 10^9 \times 0.5$ bytes \\
During computation & \multicolumn{2}{c}{$14 \times 10^9 \times 2$ bytes} \\
\hline
\textbf{Memory Consumption} & $\approx 56$ GB & $\approx 7$ GB during loading \\
\hline
\end{tabular}
\caption{\centering Memory usage before and after quantization}
\label{table:memory_use}
\end{table}

Leveraging a single GeForce RTX 4090 with 24GB of VRAM for fine-tuning a 14-billion-parameter model, we initially applied quantization to reduce the memory usage and accelerate inference, though at the potential cost of precision loss. During loading, the model was quantized to 4-bit precision and subsequently converted to 16-bit for computations. Post-loading, neither the original nor the quantized weights were retained in memory.

The fine-tuning without quantization approach utilized LoRA under the deepspeed's ZeRO-3 optimization. We employed three GeForce RTX 4090 GPUs, each with 24GB of memory, which allowed the fine-tuning of the model on a dataset of over 4,000 entries. The process took approximately seven hours to complete.

Both fine-tuning methodologies proved to be effective, demonstrating the practical applicability of our approaches to large-scale model optimization.

\newpage

\section{Chemistry related indicators and examples}\label{esi_compass}

We assessed the chemistry ability of the LLMs using the chemistry test questions from C-Eval, which comprises multiple discipline questions in multiple levels in Chinese (\autoref{esi_indicators}). This test was completed in Chinese since SynAsk's original language is Chinese. However, we acknowledge that with the LLMs' powerful language ability, testing of the LLMs with different major languages in the world would reach close results.

We provide a set of example questions for the chemistry question in C-Eval at multiple levels. Sections \autoref{subsec:college_chem}, \autoref{subsec:high_school_chem}, and \autoref{subsec:middle_school_chem} refers to the chemistry questions at college, high school and middle school levels, respectively. The dataset format consists of multiple-choice questions and answers. The Prediction contains the answers predicted by three models: SynAsk, Qwen1.5-14B-Chat, and Qwen-14B-Chat.

\subsection{C-Eval (College Chemistry)}\label{subsec:college_chem}
\begin{tcolorbox}
\textbf{Problem:} \\
\begin{CJK}{UTF8}{gbsn}
以下是中国关于大学化学考试的单项选择题，请选出其中的正确答案。\\
下列说法中，正确的是：
\begin{enumerate}[label=(\Alph*)]
\item 单质的焓为零
\item 反应的热效应就是该反应的摩尔焓变
\item 单质的摩尔生成焓为零
\item 由最稳定单质生成1 mol化合物时，该化合物的标准摩尔生成焓 $\Delta_{\mathfrak{f}}H_{\mathfrak{m}}^{\mathfrak{e}}$ 等于该生成反应的 $\Delta_{\mathfrak{r}}H_{\mathfrak{m}}^{\mathfrak{e}}$ 
\end{enumerate}
\end{CJK}
\

\textbf{English translation:} \\
The following are single-choice questions on university chemistry exams in China. Which of the following statements is correct?

\begin{enumerate}[label=(\Alph*)]
\item The enthalpy of an element is zero
\item The heat of reaction is equal to the molar enthalpy change of the reaction
\item The molar enthalpy of formation of an element is zero
\item When 1 mole of a compound is formed from the most stable elements, the standard molar enthalpy of formation $\Delta_{\mathfrak{f}}H_{\mathfrak{m}}^{\mathfrak{e}}$ of the compound is equal to the standard molar enthalpy of reaction $\Delta_{\mathfrak{r}}H_{\mathfrak{m}}^{\mathfrak{e}}$ of the formation reaction
\end{enumerate}

\

\textbf{Answer:} D 

\

\textbf{Predictions:} 
\begin{itemize}[label={}]
    \item \textbf{SynAsk:} D
    \item \textbf{Qwen1.5-14B-Chat:} C
    \item \textbf{Qwen-14B-Chat:} C
\end{itemize}
\end{tcolorbox}

\subsection{C-Eval (High School Chemistry)}\label{subsec:high_school_chem}
\begin{tcolorbox}
\begin{CJK}{UTF8}{gbsn}
\textbf{Problem:} \\
以下是中国关于高中化学考试的单项选择题，请选出其中的正确答案。\\
下列说法中，正确的是：\\
在一定温度下的恒容密闭容器中，当下列哪些物理量不再发生变化时，表明下述反应：\\
$A(s) + 2B(g) \rightleftharpoons C(g) + D(g)$ 已达到平衡状态：\\
\ding{172}混合气体的压强\\
\ding{173}混合气体的密度\\
\ding{174}B的物质的量浓度\\
\ding{175}气体的总物质的量\\
\ding{176}混合气体总质量

\begin{enumerate}[label=(\Alph*)]
\item \ding{173}\ding{174}\ding{176}
\item \ding{172}\ding{173}\ding{174}
\item \ding{173}\ding{174}\ding{175}
\item \ding{172}\ding{174}\ding{175}\ding{176}
\end{enumerate}
\end{CJK}
\

\textbf{English translation:}

The following are single-choice questions on high school chemistry exams in China. Please select the correct answer.

Which of the following statements is correct?

In a constant-volume sealed container at a certain temperature, when which of the following physical quantities no longer change, it indicates that the following reaction:
\[ A(s) + 2B(g) \rightleftharpoons C(g) + D(g) \]
has reached equilibrium:

\begin{itemize}
\item[\ding{172}] Pressure of the mixed gases
\item[\ding{173}] Density of the mixed gases
\item[\ding{174}] Concentration of substance B
\item[\ding{175}] Total amount of gas
\item[\ding{176}] Total mass of the mixed gases
\end{itemize}

\begin{enumerate}[label=(\Alph*)]
\item \ding{173}\ding{174}\ding{176}
\item \ding{172}\ding{173}\ding{174}
\item \ding{173}\ding{174}\ding{175}
\item \ding{172}\ding{174}\ding{175}\ding{176}
\end{enumerate}

\

\textbf{Answer:} A 

\

\textbf{Predictions:} 
\begin{itemize}[label={}]
    \item \textbf{SynAsk:} A
    \item \textbf{Qwen1.5-14B-Chat:} \ding{173}\ding{174}\ding{176}
    \item \textbf{Qwen-14B-Chat:} A
\end{itemize}
\end{tcolorbox}

It is noted while Qwen1.5-14B-Chat provides with the right answer, it predicts with the context of the answer directly without showing the correct choice ``A". 

\subsection{C-Eval (Middle School Chemistry)}\label{subsec:middle_school_chem}

\begin{tcolorbox}
\begin{CJK}{UTF8}{gbsn}
\textbf{Problem:} \\
以下是中国关于初中化学考试的单项选择题，请选出其中的正确答案。\\
下列有关实验现象的描述正确的是：

\begin{enumerate}[label=(\Alph*)]
\item 硫在氧气中燃烧发出淡蓝色火焰
\item 无色酚酞试液遇稀盐酸变成红色
\item 硫酸铜溶液和氢氧化钠溶液反应会产生蓝色沉淀
\item 红磷在空气中燃烧产生白雾
\end{enumerate}
\end{CJK}
\

\textbf{English translation:} \\
The following are single-choice questions on junior high school chemistry exams in China. Please select the correct answer. \\
Which of the following descriptions about experimental phenomena is correct?

\begin{enumerate}[label=(\Alph*)]
\item Sulfur burns with a pale blue flame in oxygen
\item Colorless phenolphthalein turns red when mixed with dilute hydrochloric acid
\item The reaction between copper sulfate solution and sodium hydroxide solution produces a blue precipitate
\item Red phosphorus burns in air to produce white smoke
\end{enumerate}

\

\textbf{Answer:} C 

\

\textbf{Predictions:} 
\begin{itemize}[label={}]
    \item \textbf{SynAsk:} C
    \item \textbf{Qwen1.5-14B-Chat:} C
    \item \textbf{Qwen-14B-Chat:} C
\end{itemize}
\end{tcolorbox}

\vspace{1em}

\newpage
\section{Reaction yield prediction results}\label{esi_rxn_pred}
We randomly collected 60 recent published S\textsubscript{N}Ar reactions from the last three years (2022-2024), which includes new substrate molecules and never seen by the reaction model. The model predicted yield versus literature reported yield are compared in the attached spreadsheet file, SI.xlsx, with an MAE of 14.1\%. These recent published reactions consist of seven different reaction conditions.

\pagebreak
\section{Retrosynthetic pathway of selected target molecules}\label{esi_retrosynthesis}

\begin{figure}[H]
\centering
\includegraphics[width=1\textwidth]
{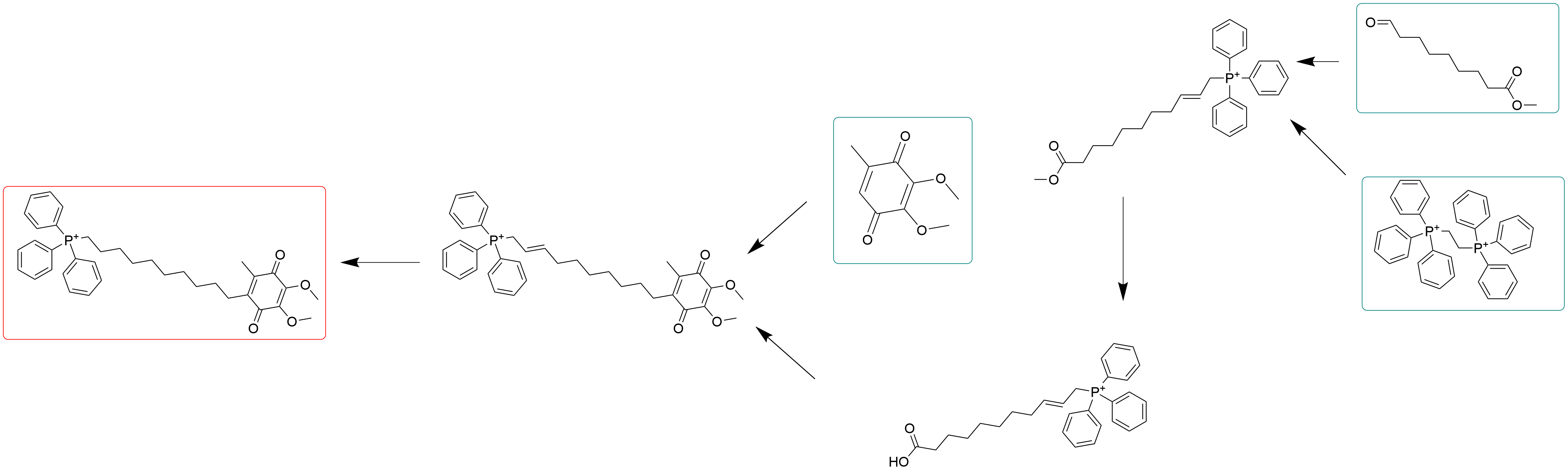}
\caption{\centering The synthetic route of the target molecule 	
mitoquinone planned by SynAsk's retrosynthetic tool.}\label{route1}
\end{figure}

\begin{figure}[H]
\centering
\includegraphics[width=1\textwidth]
{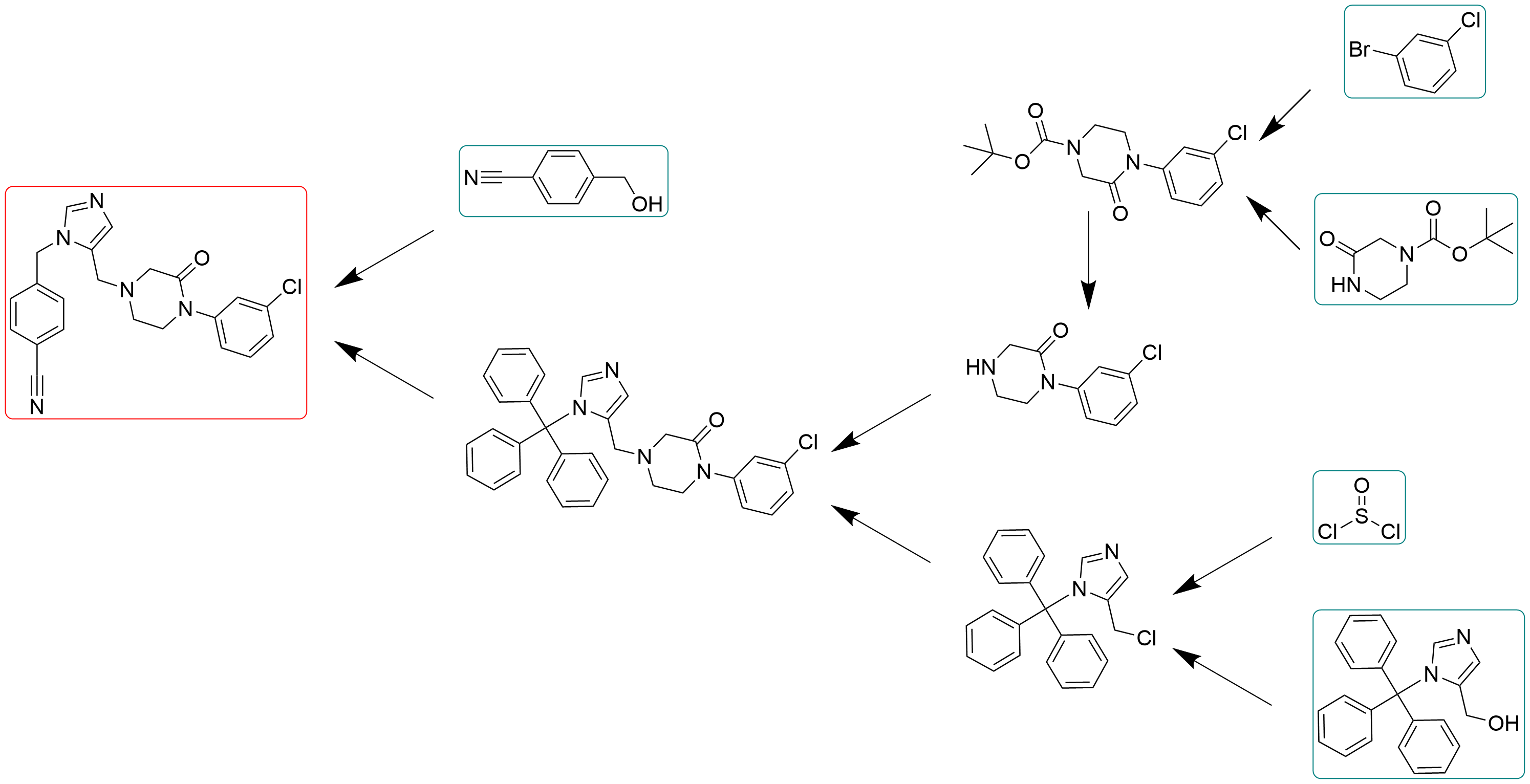}
\caption{\centering The synthetic route of the target molecule L-778123 planned by SynAsk's retrosynthetic tool.}\label{route2}
\end{figure}

\begin{figure}[H]
\centering
\includegraphics[width=1\textwidth]
{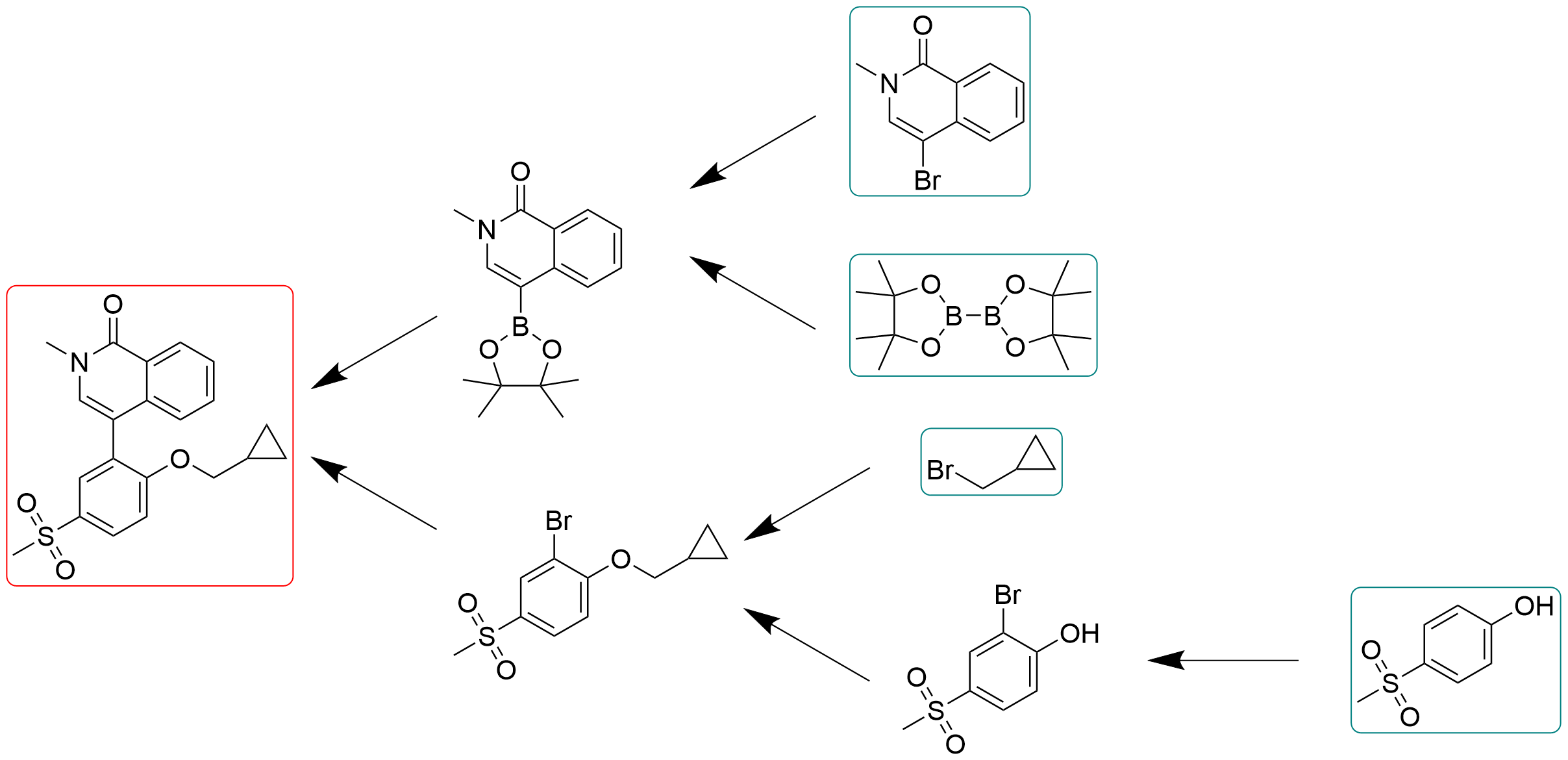}
\caption{\centering The synthetic route of the target molecule trotabresib planned by SynAsk's retrosynthetic tool.}\label{route3}
\end{figure}

\begin{figure}[H]
\centering
\includegraphics[width=1\textwidth]
{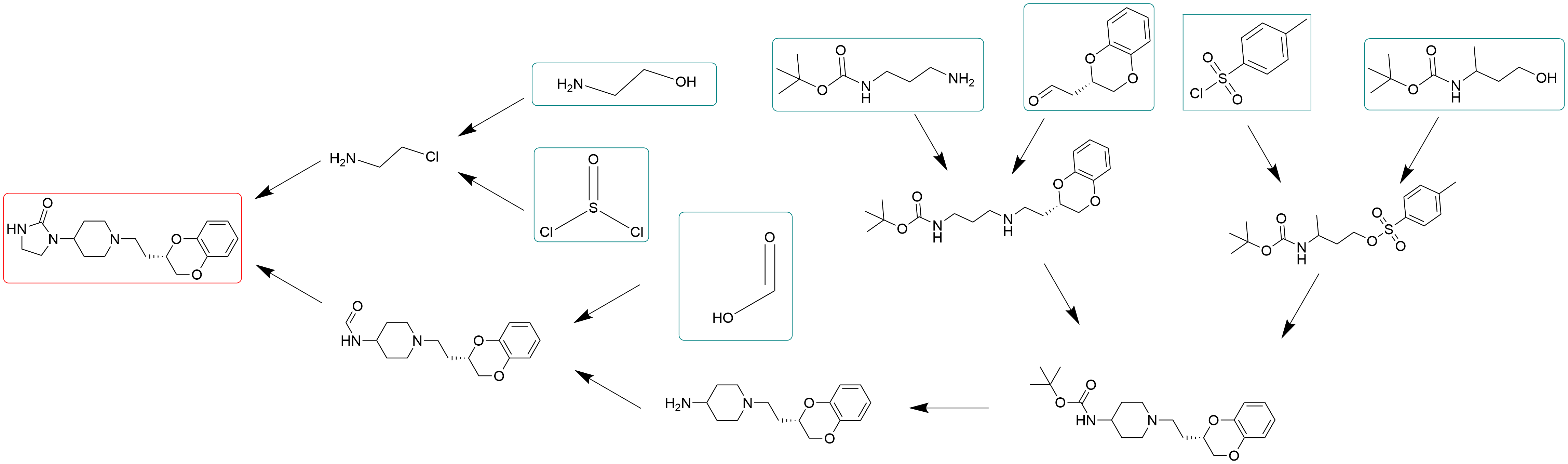}
\caption{\centering The synthetic route of the target molecule 	
azaloxan planned by SynAsk's retrosynthetic tool.}\label{route4}
\end{figure}

\autoref{route1}, \autoref{route2}, \autoref{route3} and \autoref{route4} shows numbers of retrosynthetic pathways generated by SynAsk, which provide insights for synthetic chemists. The routes indicate the ability of SynAsk in computer assisted synthetic planning (CASP). 

We are developing strategies towards generation of more reasonable retrosynthetic pathways. This will be published elsewhere, and integrated into SynAsk. Till now, no efforts were made to experimentally validate the synthetic routes provided in the ESI, and more synthetic routes to other target molecules can be generated via command to SynAsk.

\section{Examples of the SynAsk platform outputs versus other LLMs}\label{esi_llm_example}
\setcounter{figure}{4}
\begin{figure}[H]
\centering
\includegraphics[width=1\textwidth]
{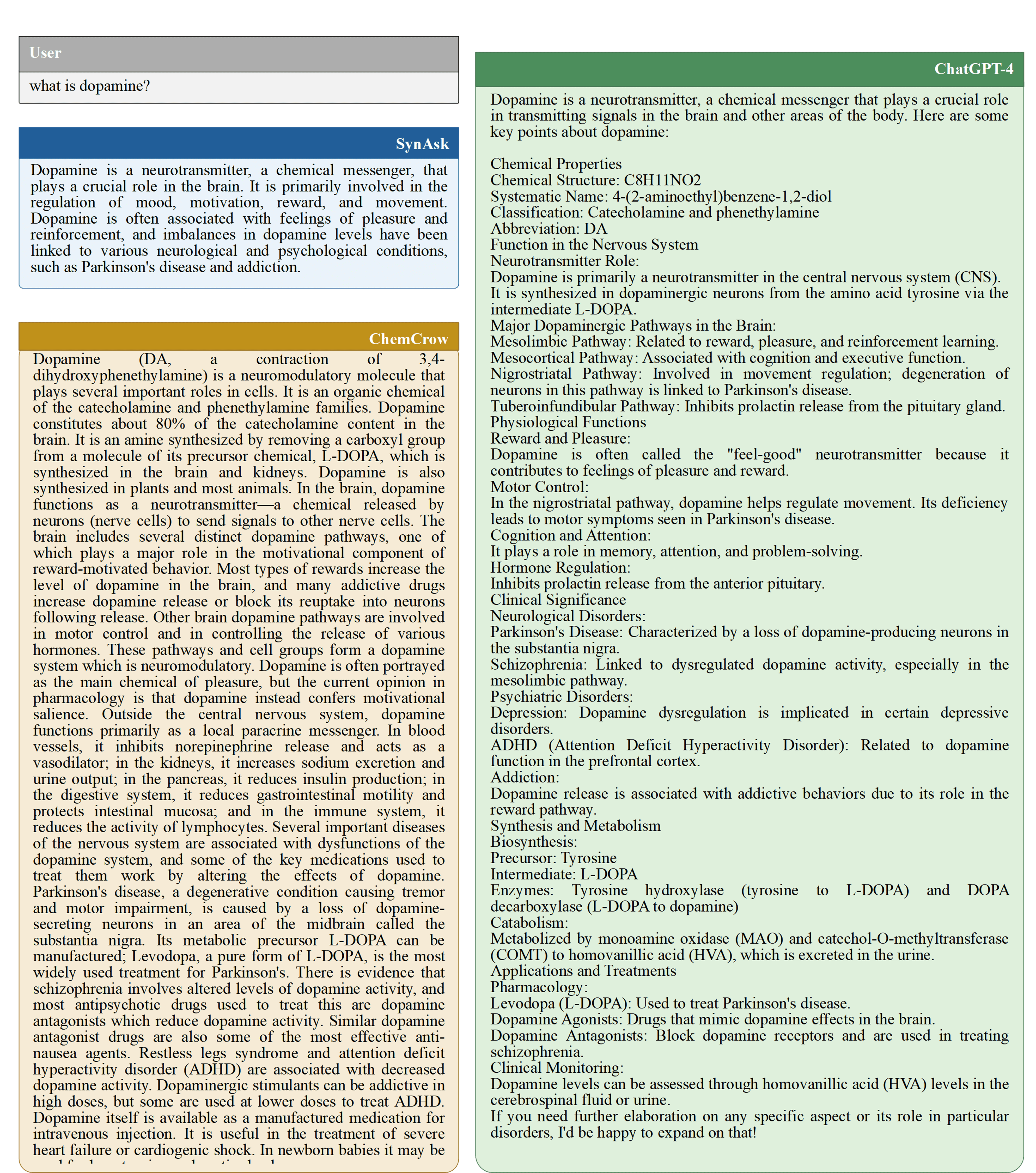}
\caption{\centering The first example of the outputs from the LLMs.}\label{example1}
\end{figure}

\begin{figure}[H]
\centering
\includegraphics[width=1\textwidth]
{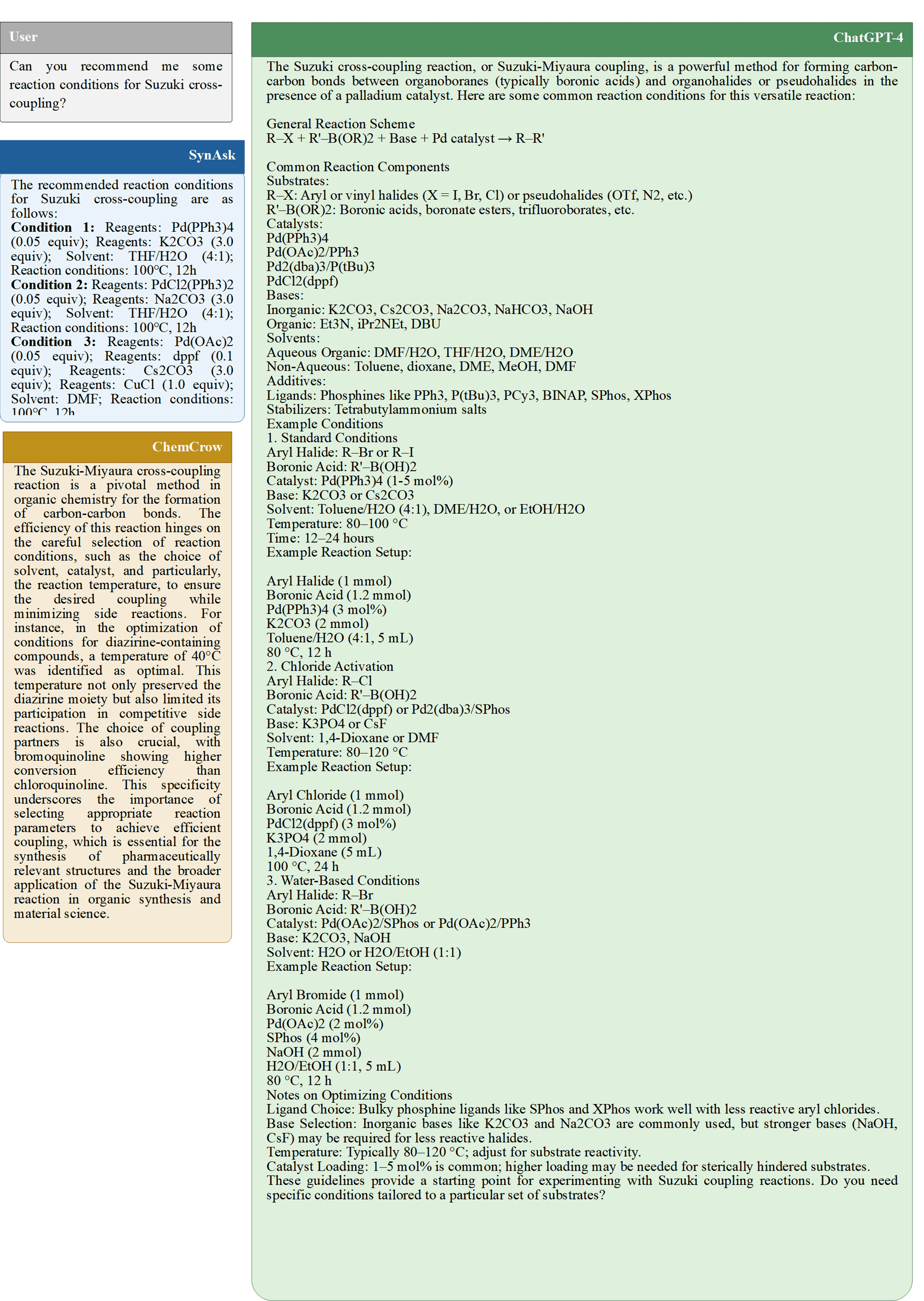}
\caption{\centering The second example of the outputs from the LLMs.}\label{example2}
\end{figure}

\begin{figure}[H]
\centering
\includegraphics[width=1\textwidth]
{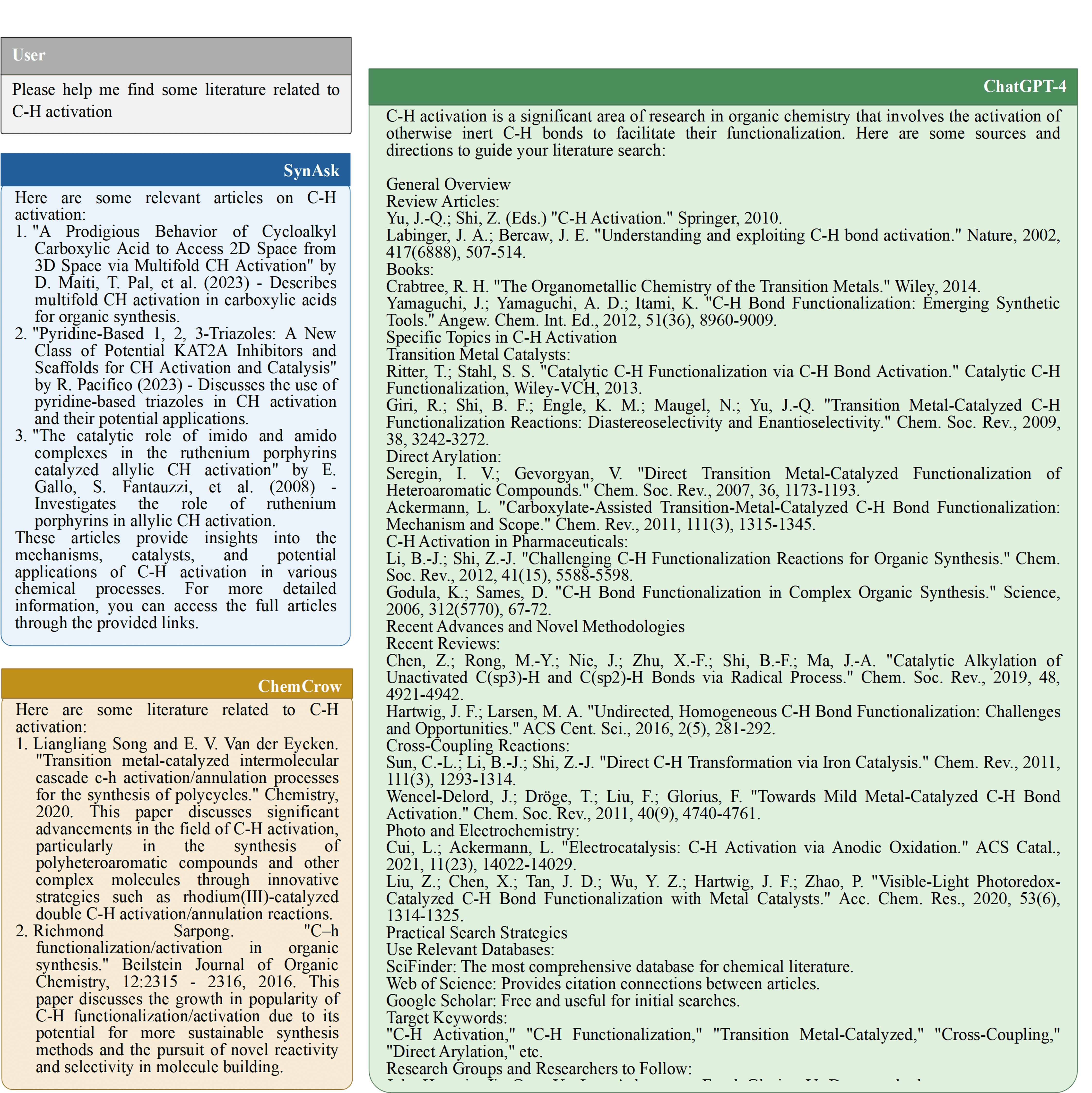}
\caption{\centering The third example of the outputs from the LLMs.}\label{example3}
\end{figure}

\end{document}